\documentstyle[epsfig,12pt]{article}

\begin{document}

\begin{center}

\title "Estimation of steady-state basic parameters of stars

\bigskip
\author "B.V.Vasiliev
\bigskip

Institute in Physical and Technical Problems, Dubna, Russia,
141980
\bigskip

$e-mail: vasiliev@main1.jinr.ru$
\end{center}

\bigskip
\begin{abstract}

From a minimum of total energy of
celestial bodies, their basic parameters are obtained. The
steady-state values of mass, radius, and temperature of stars and
white dwarfs, as well as masses of pulsars are calculated. The
luminosity and giromagnetic ratio of celestial bodies are
estimated. All the obtained values are in a satisfactory agreement
with observation data.

\end{abstract}



\section{Introduction}

It is agreed-upon to consider that pressure and density of a
substance inside stars grow more or less monotonously approaching
their centrs. It is supposed even that an ultrahigh-density
substance can exist near the center of some stars. Thus, it is
considered that the basic parameters of stars, such as mass,
radius and temperature, are  more or less arbitrary.

However, density, mass, and temperature of a star can be
determined coming from the requirement of a minimum of energy if a
star exists in an equilibrium state.

The effect of gravity-induced electric polarization of plasma
inside a star \cite {1} gives a basis for such a calculation.

Under the influence of gravity,  the nuclei of plasma "hang" on
their electronic clouds. Therefore, they are always located right
below these clouds. As a result, electrons form a stratum at the
external surface of a plasma body, and inside a star each cell of
plasma obtains a positive charge $ \delta q $. This charge is very
small and equal to $ \frac {a_0} {R_{st}} e\simeq10 ^ {-19} e $ by
the order of magnitude ($a_0 $ is Bohr radius, $R_{st}$ is the
radius of a star, and $e$ is the charge of an electron). However,
it is sufficient, that at each point inside a star a gravitational
force is cancelled by an electric force

\begin{equation}
m_i\vec{g}+ \delta q \vec{E}=0,\label{1010}
\end{equation}
\noindent
where $m_i $ is the mass of an ion, $ \vec {g} $ is gravity
acceleration, $ \vec {E} $ is an electric field intensity created
by charges $ \delta q $

\begin{equation}
div \vec{E}= 4\pi \delta q n, \label{1020}
\end{equation}
\noindent
where $n$ is the electron density.

From here it follows that the gradient of pressure inside a star
is equal to zero, and the substance has a steady value of density.
For a star consisting of degenerate relativistic plasma, the
temperature effects can be neglected. For a star consisting of hot
nonrelativistic plasma, the absence of the gradient of pressure
and density of a substance requires an assumption that temperature
inside a star is constant. It concerns, certainly, only the basic
core of stars consisting of electrically polarized plasma. There
is a thin stratum at the surface of a star, where a substance
exists in an usual atomic state. In this stratum, the gradient of
pressure, the gradient of density, and the gradient of temperature
are present. Because of a small thickness of this stratum (about
$10^{-2} R_{st}$), its role in the energy balance of the whole
star can be neglected.

Since the gravitational force is cancelled by the electric force
in cores of stars, the gravitational energy is compensated by the
electric energy inside a core as well \cite{1} (the gravitational
energy is negative). There is an uncompensated part of energy of a
gravitational field. It is the energy of gravitational field
outside a star

\begin{equation}
E_G=-\frac{GM^2}{2R},\label{1040}
\end{equation}
where $G$ is the gravitational constant, $M$ and $R$ are mass and
radius of a star. Because plasma is electroneutral as a whole, one
proton should be related to electron of the Fermi gas of plasma.
The existence of one neutron per proton is characteristic of a
substance consisting of light nuclei. The quantity of neutrons
increases approximately to 1.8 per proton for the heavy-nuclei
substance, i.e. to one electron in plasma a mass equal to $m_p
\biggl (\frac {A} {Z} \biggr)$ should be related. Thus

\begin{equation}
M=N m_p \biggl (\frac {A} {Z} \biggr).\label{1050}
\end{equation}
Here $N$ is a full number of electrons in a star, $m_p$ is the
proton mass, $A$ and $Z$ are the mass number and charge of a
nucleus, respectively. For hydrogen $ \biggl (\frac {A} {Z}
\biggr) =1 $. Therefore,

\begin{equation}
1\leq \biggl(\frac{A}{Z}\biggr)<2.8.\label{1060}
\end{equation}
The field Eq.({\ref{1040}}) tends to compress a star as a whole.

The jump in electric polarization at the surface of a star core
produces additional pressure inside a star \cite {1}, since the
jump in polarization is accompanied  by the pressure jump. As this
energy is considerably smaller than the gravitational one, it can
be neglected for simplicity.

At high temperature the energy of the black radiation can be
important in a general balance of the star energy

\begin{equation}
E_r=\frac{4\sigma}{c}T^4 V\label{1070}
\end{equation}
where $\sigma=5.7\cdot 10^{-5}$ $\frac{g}{cm^3 K^4}$ is the
Stefan-Boltzmann constant, and $V$ is the volume of a star.

Thus, the total energy of a star is a sum of the total energy of
plasma and the energy of black radiation

\begin{equation}
E_{total}=E_{pl}+E_{r}.\label{1080}
\end{equation}

\section {A star consisting of hot nonrelativistic electron - nuclear plasma}

The energy of plasma in a star is a sum of its potential energy
$U$ and kinetic energy $E_{kinetic}$. According to the virial
theorem the potential energy of a system of interacting charged
particles and their kinetic energy are connected \cite {2,3}

\begin{equation}
U=-2E_{kinetic}.\label{2010}
\end{equation}
Because the ion mass is large, the  chemical potential of ions is
small and their kinetic energy can be neglected. Thus, electrons
will make a main contribution into the kinetic energy of plasma.

We name plasma a hot one, if the temperature inside a star is more
then the degeneration temperature of the electron gas and the
electric interaction between particles of plasma can be neglected

\begin{equation}
\biggl(\frac{kT}{e^2}\biggr)^3>>n.\label{2020}
\end{equation}
Under this condition it is possible to consider the electron gas
of plasma an ideal one, and write its equation of state as the
ideal gas law

\begin{equation}
PV = NkT.\label{2030}
\end{equation}
Thus, the kinetic energy of plasma in a star

\begin{equation}
E_{kinetic}=\frac{3}{2}NkT,\label{2040}
\end{equation}
where $k$ is the Boltzmann constant.

Because gravity  does not influence photons and acts on heavy ions
only, the total energy of hot plasma is

\begin{equation}
E_{pl}=-\frac{GM^2}{2R}+\frac{3}{2}NkT,\label{2050}
\end{equation}

or, according to the virial theorem (Eq.({\ref{2010}})), the
energy of plasma in a star

\begin{equation}
E_{pl}=-\frac{GM^2}{4R}=-\frac{3}{2}NkT\label{2060}
\end{equation}

and the total energy of a hot star is

\begin{equation}
E_{total}=-\frac{GM^2}{2R}+\frac{3}{2}NkT+\frac{4\sigma}{c}T^4 V =
-\frac{GM^2}{4R}+\frac{4\sigma}{c}T^4 V \label{2070}
\end{equation}

\section {Calculation of the corrections to the electron energy}

This equation is valid for an electron gas at very high
temperature, when the identity of particles is of no significance.
At finite temperature the identity of particles leads to an extra
stiffness of the electron gas than it follows from Eq.({\ref
{2030}}). On the other hand, electron gas in plasma will exhibit
more softness under the influence of a Coulomb field of nuclei. We
shall carry out the account of these corrections according to the
methods described by Landau and Lifshits \cite{2}.

\subsection {The correction on electron identity}

The total energy of the electron gas is known \cite {2}

\begin{equation}
E_e=\int_0^\infty \varepsilon dN_\varepsilon= \frac{2^{1/2}V
m^{3/2}}{\pi^2 \hbar^3}
\int_0^\infty\frac{\varepsilon^{3/2}d\varepsilon}
{e^{(\varepsilon-\mu)/kT}+1}, \label{2110}
\end{equation}
where $\varepsilon=P^2/{2m}$, $P$ and $m$ are momentum and mass of
an electron in the nonrelativistic gas.

According to the definition of a chemical potential of electrons
\cite{2},

\begin{equation}
\mu= kT ln \biggl[\frac{N}{2V}\biggl(\frac{2\pi \hbar^2}{m
kT}\biggr)^{3/2}\biggr],\label{2120}
\end{equation}
and at high temperature

\begin{equation}
e^{\mu/kT}<<1.\label{2130}
\end{equation}

Therefore, in this case the integral in Eq.({\ref {2110}}) can be
expanded into the power series $e^{\mu/kT-\varepsilon/kT}$.

Keeping the first terms of series, an approximated expression of
the free energy of the hot electron gas with the account of the
correction on the identity of electrons is obtained

\begin{equation}
F=F_{ideal}+N\frac{\pi^{3/2}e^3 a_0^{3/2}n}{4(kT)^{1/2}}.
\label{2160}
\end{equation}

\subsection {The estimation of the influence of nuclei on the
hot electron gas}

The nuclei reduce pressure of electron gas apart from them,
therefore, this correction to pressure must be negative. The
estimation of this correlative correction can be carried out by
the method offered by Debye and H\"{u}ckel \cite {2}.

The energy of an charged particle inside plasma is equal to
$e\varphi_e $, where $e$ is a charge of a particle, and $
\varphi_e $ is the electric potential created by other particles
on this particle. This potential in plasma is determined by the
Debye law

\begin{equation}
\varphi=\frac{e}{r}e^{-\frac{r}{r_D}}.\label{3010}
\end{equation}
As in dense plasma nuclei form an ordered lattice \cite{2},
electrons play a role in screening charges only.

Thus, the Debye radius is

\begin{equation}
r_D=\sqrt{\frac{kT}{4\pi e^2 n}}. \label{3020}
\end{equation}
At the small value of ratio $\frac {r} {r_D}$, the potential can
be expanded into the series

\begin{equation}
\varphi=\frac{e}{r}-\frac{e}{r_D} + ...\label{3030}
\end{equation}
The following terms are converted into zero at $r=0$. The first
term of this series is the potential of the electron. The second
term is the potential created by other particles on electron.
Therefore, the energy of the electron gas in plasma created by the
action of screened fields of other particle:

\begin{equation}
E_{corr}=-\frac{e^2}{r_D}N=2e^3\sqrt{\frac{\pi}{kT}}\frac{N^{3/2}}{V^{1/2}}.\label{3040}
\end{equation}
By using of the known Gibbs-Helmholtz equation \cite {2}

\begin{equation}
\frac {E} {T^2} = -\frac {\partial} {\partial T} \frac {F} {T},
\label{3050}
\end{equation}
we obtain the correction to the free energy

\begin{equation}
F=F_{ideal} -N\frac{2 e^3}{3}\sqrt{\frac{\pi n}{kT}}.\label{3060}
\end{equation}

\subsection {The steady-state value of a star density}

Accounting for these corrections on electron identity and
influence of ions, the free energy of the hot electron gas
acquires the form

\begin{equation}
F=F_{ideal}+N\frac{\pi^{3/2}e^3 a_0^{3/2}}{4(kT)^{1/2}}n -
N\frac{2 e^3}{3}\frac{\pi^{1/2}}{(kT)^{1/2}}n^{1/2}\label{4010}
\end{equation}
From the equilibrium condition

\begin{equation}
\biggl(\frac{\partial F}{\partial n}\biggr)_{N,T}=0,\label{4020}
\end{equation}
we obtain a steady-state value of electron density in plasma of a
star

\begin{equation}
n_{st}=\frac{16}{9 \pi^2 a_0^3}\simeq 2\cdot 10^{23}
cm^{-3}.\label{4030}
\end{equation}
For a star with a sufficiently high temperature, according to
Eq.({\ref {2020}}), it is energetically favourable to have this
density.

\subsection {The steady-state values of the mass, the radius and temperature of a star}

According to Eq.({\ref {2070}}), the total energy of a star

\begin{equation}
E_{total}\cong -\frac{3}{2}k T n_{st} V+\frac{4 \sigma} {c} T^4
V.\label{4040}
\end{equation}
This function has a minimum at the temperature

\begin{equation}
T_{st}=\biggl(\frac{3 ck n_{st}}{32 \sigma}\biggr)^{1/3}
=\biggl(\frac{c k}{6 \pi^2 \sigma a_0^3} \biggr)^{1/3} \simeq
2.5\cdot 10^7 K.\label{4070}
\end{equation}
According to the virial theorem Eq.({\ref{2010}}), at an
equilibrium state into the electron gas subsystem

\begin{equation}
\frac{GM^2}{2R}=3NkT.\label{4080}
\end{equation}
With the account of Eq.({\ref {4030}}) and Eq.({\ref {4070}}), we
can calculate a steady-state value of radius of hot stars

\begin{equation}
R_{st}=\biggl[\biggl(\frac{3}{2}\biggr)^2 \biggl(\frac{\pi c
k^4}{48\sigma G^3 m_p^6}\biggr)^{1/6}\biggr]
\frac{a_0}{\bigl(\frac{A}{Z}\bigr)}\simeq
\frac{1.6}{\bigl(\frac{A}{Z}\bigr)} R_\odot \label{4090}
\end{equation}
where $R_\odot $ is the radius of the Sun.

The steady-state value of mass of hot stars

\begin{equation}
M_{st}\simeq \biggl[\frac{27}{4}\biggl(\frac{c}{3\pi\sigma G^3
}\biggr)^{1/2}
\biggl(\frac{k}{m_p^{3/2}}\biggr)^2\biggr]\frac{m_p}
{\bigl(\frac{A}{Z}\bigr)^{2}}\simeq
\frac{1.1}{\bigl(\frac{A}{Z}\bigr)^2} M_\odot, \label{4100}
\end{equation}
where $M_\odot $ is the mass of the Sun.

\section {The equilibrium condition of a white dwarf}

    When density of a substance is sufficiently high and temperature is
insignificant, the electrons form a degenerate and relativistic
Fermi gas. Thus, a star obtains the other equilibrium state.

It is known \cite {2} that the electron gas is a relativistic one,
when its Fermi momentum

\begin {equation}
p_F = (3\pi^2) ^ {1/3} n ^ {1/3} \hbar > mc,\label {5010}
\end {equation}
i.e., at density of particles $n>10^{31}cm^{-3}$, which is
characteristic for white dwarfs.

For a degeneration, electron gas temperature must be

\begin {equation}
T << \frac {mc^2} {k} \approx 10 ^ {10} K.\label {5020}
\end {equation}
The energy of the relativistic electron gas is known \cite {2}

\begin {equation}
 E=\frac{3}{4}\biggl(3\pi^2\biggr)^{1/3}\hbar c N
 \biggl(\frac{N}{V}\biggr)^{1/3}.\label{5030}
\end {equation}
Therefore, the total energy of a white dwarf

\begin {equation}
 E_{total}=-\frac{GM^2}{2R}+\frac{3}{4}\biggl(3\pi^2\biggr)^{1/3}\hbar c N
 \biggl(\frac{N}{V}\biggr)^{1/3}.\label{5040}
\end {equation}
As temperature is defined by the inequality Eq.({\ref {5020}}),
the temperature depending terms can be neglected. It is possible
under the condition

\begin {equation}
\frac{4\sigma}{c}T^4 V <<\frac{3}{4}\bigl(3\pi^2\bigr)^{1/3}\hbar
c N \biggl(\frac{N}{V}\biggr)^{1/3}.\label{5045}
\end {equation}
As according to the virial theorem (Eq.({\ref {2010}}))

\begin {equation}
 \frac{GM^2}{2R}=\frac{3}{2}\biggl(3\pi^2\biggr)^{1/3}\hbar c N
 \biggl(\frac{N}{V}\biggr)^{1/3},\label{5050}
\end {equation}
the full number of electrons inside any white dwarf is fixed

\begin {equation}
N_{dw} =\frac{M}{\bigl(\frac{A}{Z}\bigr)
m_p}=\frac{3^{5/2}\pi^{1/2}}{2} {\biggl (\frac {\hbar c} {G
\bigl(\frac{A}{Z}\bigr)^2 m_p^2} \biggr)^{3/2}}\simeq \frac
{3.2\cdot 10 ^ {58}} {\bigl (\frac {A} {Z} \bigr) ^3} .\label
{5060}
\end {equation}
Thus, in the equilibrium state any white dwarf should have the
steady-state value of mass, depending on world constants only and
the ratio $ \biggl (\frac {A} {Z} \biggr) $:

\begin {equation}
M_{dw} =\biggl[\frac {3^{5/2}\pi^{1/2}}{2} \biggl (\frac {\hbar c}
{G m_p^2}\biggr)^{3/2}\biggr]
\frac{m_p}{\bigl(\frac{A}{Z}\bigr)^2}\simeq
\frac{12.5M_\odot}{{\bigl(\frac{A}{Z}\bigr)^2}}.\label{5070}
\end {equation}
According to the inequality Eq.({\ref {5010}}), the steady-state
value of radius of a white dwarf

\begin {equation}
R_{dw}<\biggl(\frac{9\pi}{4}N_{dw}\biggr)^{1/3}\frac{\hbar}{mc}
\simeq 3\cdot 10^{-2}R_\odot \label{5080}
\end {equation}
and the density of electrons

\begin {equation}
n_{dw}=\frac{N_{dw}}{\frac{4\pi}{3}R_{dw}^3}>>10^{30}cm^{-3}.\label{5090}
\end {equation}
This correlates well with Eq.({\ref {5010}}).

According to Eq.({\ref {5045}})

\begin {equation}
T_{dw}<<\frac{m
c^{3/2}}{(4\pi)^{1/2}\sigma^{1/4}\hbar^{3/4}}\simeq 3\cdot
10^{9}K.\label{5100}
\end {equation}
Which is consistent with Eq.({\ref {5020}}).

The steady-state value of mass of a star ({\ref {5070}}),
consisting from the relativistic electron gas, does not depend on
mass of an electron. Due to this fact, this equation is valid for
stars, consisting of a degenerate gas of other relativistic Fermi
particles, in particular, neutrons. Therefore, if we can consider
a pulsar as a neutron star containing a small number of protons
and electrons (with concentration more than $10^{-19}$), the
equation Eq.({\ref {5070}}) should determine the steady-state
value of mass of a pulsar.

\section {The mass-luminosity relationship}

It seems possible to assume that  temperature inside a star
determines temperature at its surface. If the relation between a
surface temperature $T$ and the  steady-state values of inner
temperature $T_{st}$ is constant for all stars

\begin {equation}
T = const\cdot T_{st},\label {6020}
\end {equation}
it is possible to deduce a relationship between a visible
luminosity of a star and its other parameters, in particular,
mass.

As the luminosity of a star

\begin {equation}
L=4\pi R^2 \frac {\alpha} {n} T^4,\label {6030}
\end {equation}
using the obtained above expressions and after simple
transformations, we obtain

\begin {equation}
 \frac{L}{L_\odot}=\biggl(\frac{M}{M_\odot}\biggr)^{10/3},\label{6040}
\end {equation}
where $L_\odot $ is the luminosity of the Sun.

\section {The comparison of the calculated values with the observation data}

The masses of stars can be measured with a considerable accuracy,
if these stars compose a binary system. There are almost 300
double stars which masses are known with the required accuracy
\cite{5}. Among these stars there are stars of the main sequence,
dwarfs, and pulsars. Approximately one half of them are visual
binaries. Their masses are measured with the a high precision.
Other half consists of spectroscopic binaries and eclipsing
binaries. For these stars the accuracy of mass measurement is a
slightly worse. Nevertheless, we shell carry out a comparison of
the calculated parameters of stars on the basis of all these
measurements.

\subsection {Masses of celestial bodies}

According to these data, the distribution of masses of hot stars
is described by the equality

\begin {equation}
<M> = (2.98\pm 0.25) M_\odot. \label {7010}
\end {equation}
It is in satisfactory agreements with the calculated steady-state
value of mass of a star Eq.({\ref {4100}}) by the order of
magnitude. Graphically this distribution is shown in Fig.1.
According to the binary star tables \cite {5}, the distribution of
white dwarfs masses is described by

\begin {equation}
<M> = (0.96\pm 0.05) M_\odot. \label {7020}
\end {equation}
This distribution is shown in Fig. 2. In this figure,
the distribution of pulsar masses \cite {6} is shown. It is described
by

\begin {equation}
<M> = (1.40\pm 0.02) M_\odot. \label {7030}
\end {equation}
It is possible to conclude that with the account of the
corrections on the factor $ \biggl (\frac {A} {Z} \biggr) $, the
results of calculations of the steady-state values of masses  are
in agreement with measured data.

\subsection {The mass-luminosity relationship}

The luminosity of stars from binary systems, depending on their
masses are shown in Fig. 3. The results of more precise
measurements of visual binaries are marked by squares. The data
for spectroscopic and eclipsing binaries are marked by triangles
and points, respectively. The line corresponds to the calculated
dependence Eq.({\ref {6040}}). It can be seen from this figure,
that in the logarithmic scale the calculated dependence is in
satisfactory agreement with the observed data.

\section {The gyromagnetic ratio of stars}

The effect of gravity-induced electric polarization \cite {1} is
characteristic for all celestial bodies, consisting of a substance
in a plasma state. The value of this polarization does not depend
on plasma properties:  no matter if it is a relativistic one or
not and if it has a degeneration or not. As $div \vec {E} =4\pi
\rho $ and $div \vec {g} = -4\pi G \gamma $ and according to
Eq.({\ref {1010}}), the gravity-induced density of the volume
electrical charge $ \rho $ is related to the density of substance
$ \gamma $ by the relation \cite {1}:

\begin {equation}
\rho =\sqrt {G}\gamma. \label {8010}
\end {equation}
Therefore a rotation of a celestial body about its axis must
induce a magnetic field. If one assumes that the electrically
polarized core of a body occupies its entire volume and if one
neglects the existence of a surface stratum, where the substance
is in an atomic state, that the gyromagnetic ratio (the relation
of a magnetic moment of a body to its angular momentum) will
obtain a steady value \cite {1}:

\begin {equation}
 \vartheta=\frac{\sqrt{G}}{3c}.\label{8020}
\end {equation}
It is possible to consider the above assumption is quite
acceptable for large celestial bodies, such as stars, and less
acceptable for planets, where the mantle can be rather large.
However, the detailed calculation for the Earth gives the result
which is in agreement with the measured value with in the accuracy
of the factor of two \cite {4}. The values of giromagnetic ratio
for all celestial bodies (for which they are known today) are
shown in Fig.4.

The data for planets are taken from \cite{7}, the data for stars
are taken from \cite{8}, and for pulsars - from \cite{9}.
Therefore, for all celestial bodies - for planets and their
satellites, for $Ap$-stars and several pulsars - the calculated
value Eq.({\ref {8020}}) with a logarithmic precision quite
satisfactorily agrees to measurements, when moments itself change
within the limits of more than 20 orders.

\begin{figure}
\begin{center}
\includegraphics[3cm,12cm][15cm,2cm]{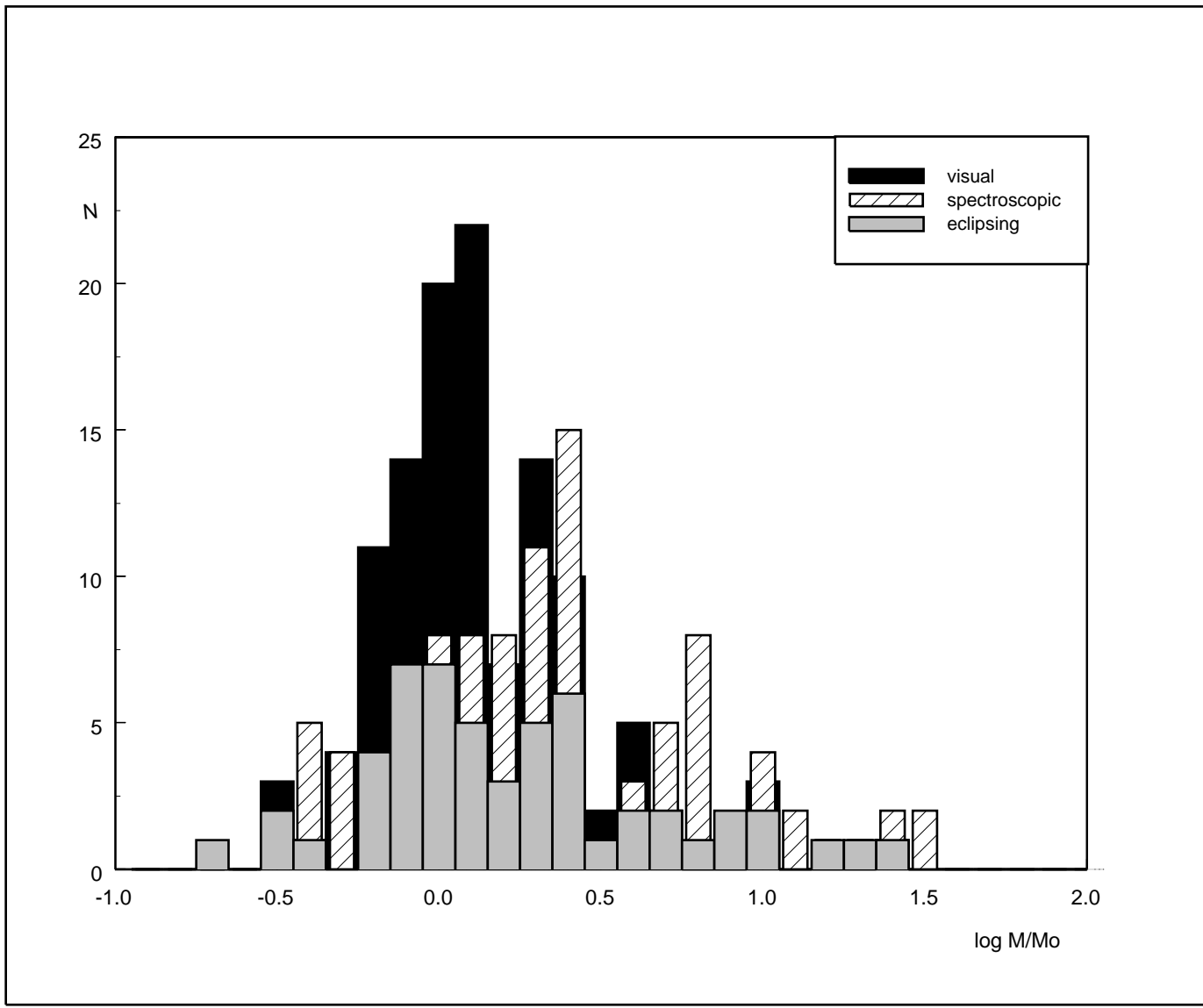}
\vspace{11cm} \caption{ The mass distribution of stars from the
binary systems \cite{5}. Data for visual, spectroscopic, and
eclipsing binaries are shown separately. On the abscissa, the
logarithm of the star mass over the Sun mass is shown. }
\label{fig1}
\end{center}
\end{figure}
\clearpage

\begin{figure}
\begin{center}
\includegraphics[3cm,12cm][15cm,2cm]{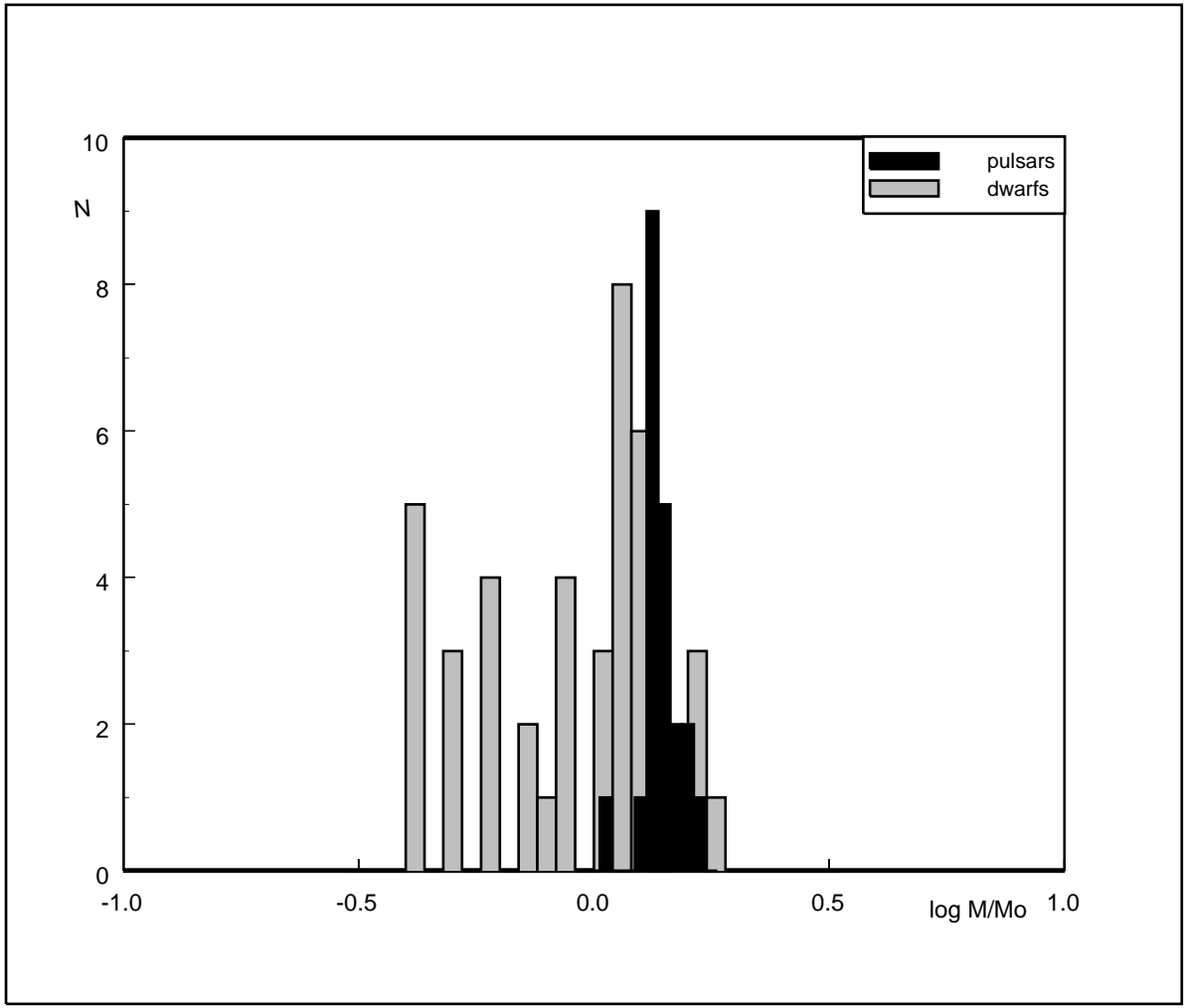}
\vspace{11cm} \caption { The mass distribution of white dwarfs and
pulsars from the binary systems \cite{5,6}. On the abscissa, the
logarithm of the star mass over the Sun mass is shown. }
\label{fig2}
\end{center}
\end{figure}
\clearpage

\begin{figure}
\begin{center}
\includegraphics[3cm,12cm][15cm,2cm]{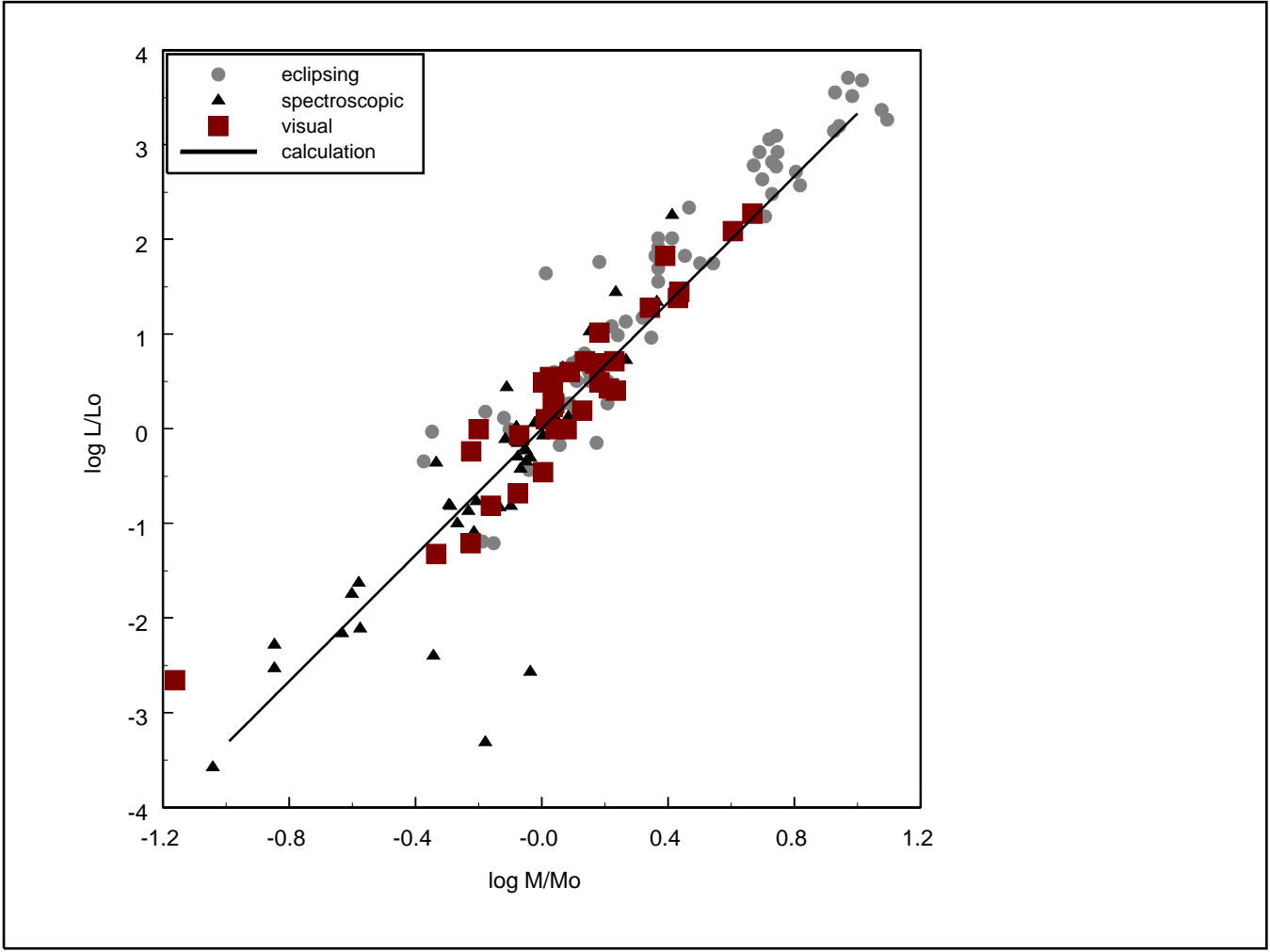}
\vspace{11cm} \caption { The luminosity of stars from binary
systems, depending on their masses are shown. The results of more
precise measurements of visual binaries are marked by squares. The
data for spectroscopic and eclipsing binaries are marked by
triangles and points, respectively. The line corresponds to the
calculated dependence Eq.({\ref {6040}}). On the ordinate, the
logarithm of the star luminosity over the Sun luminosity is shown.
On the abscissa, the logarithm of the star mass over the Sun mass
is shown. }\label{fig3}
\end{center}
\end{figure}
\clearpage

\begin{figure}
\begin{center}
\includegraphics[5cm,14cm][17cm,2cm]{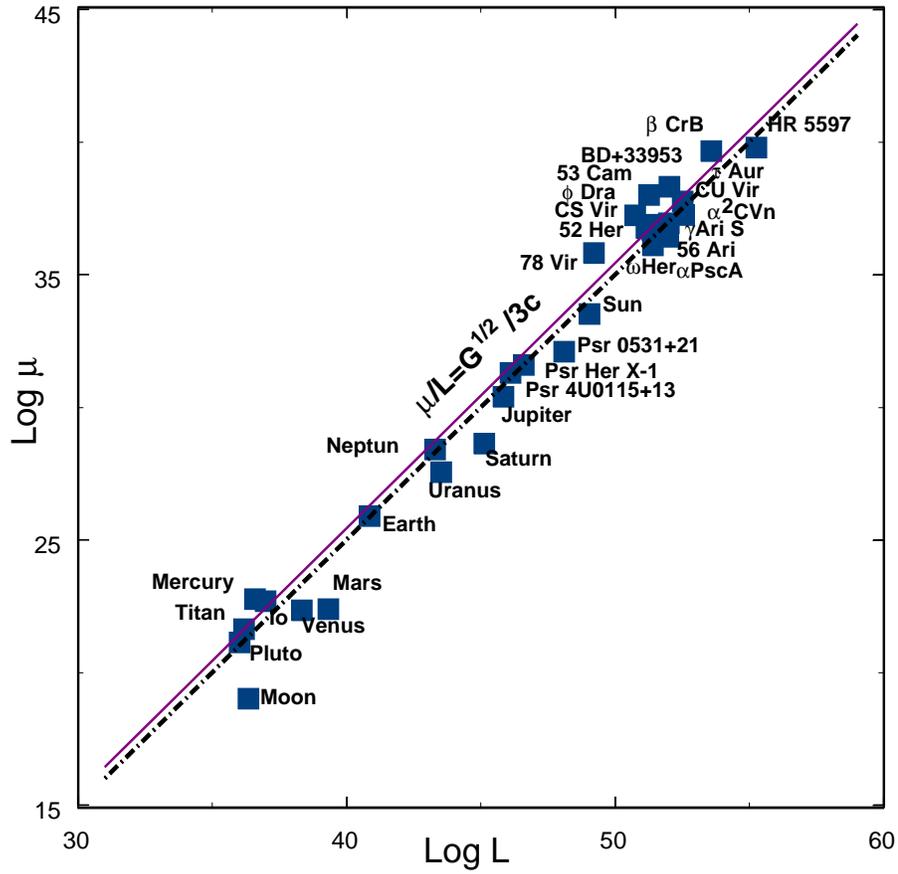}
\vspace{11cm} \caption { The observed values of the magnetic
moments of celestial bodies vs. their angular momenta. On the
ordinate, the logarithm of the magnetic moment over
$Gs\cdot{cm^3}$ is plotted; on the abscissa the logarithm of the
angular momentum over $erg\cdot{s}$ is shown. The solid line
illustrates Eq.({\ref{8020}}). The dash-dotted line is the fitting
of the observed values. } \label{fig4}
\end{center}
\end{figure}

\end{document}